\documentclass[
reprint,
longbibliography,
aps,
twoside,
bibnotes,
floatfix
]{revtex4-2}

\usepackage{physics}
\usepackage{dsfont}
\usepackage{xcolor,newfloat}
\usepackage{graphicx}
\usepackage{dcolumn}
\usepackage{bm}
\usepackage{amsmath,amssymb,amsthm,amsfonts}
\usepackage{mathtools}
\usepackage{multirow}
\usepackage{hyperref}
\usepackage[all]{hypcap}
\usepackage{chngcntr}
\usepackage{enumerate}
\usepackage{nicefrac, xfrac}

\begin{document}

\title{Boosted Entanglement Swapping}

\author{Matthias J. Bayerbach}
\thanks{These authors contributed equally to this work.}

\author{Simone E. D'Aurelio}
\thanks{These authors contributed equally to this work.}

\author{Stefanie Barz}

\affiliation{Institute for Quantum Information and Technology, University of Stuttgart, 70569 Stuttgart, Germany}
\affiliation{Center for Integrated Quantum Science and Technology (IQST), University of Stuttgart, 70569 Stuttgart, Germany}

%%%%%%%%%%%%%%%%%%%%%%%%%%%%%%%%%%
%%%%%%%%     Abstract    %%%%%%%%%
%%%%%%%%%%%%%%%%%%%%%%%%%%%%%%%%%%

\begin{abstract}
Entanglement swapping entangles photons that never interacted directly. When implemented with linear optics, its success probability is upper bounded by $50\%$, due to the limited efficiency of linear-optical Bell-state measurements. Here, we demonstrate boosted entanglement swapping using an ancilla-assisted linear-optical Bell-state measurement, which raises the theoretical success probability to $62.5\%$ and allows all four Bell states to be identified. We measure an acceptance rate of $p_{\mathrm{a}}^{\mathrm{b}}=(69.39\pm0.39)\%$ with an average fidelity of the swapped states of $\langle F^{\mathrm{b}}\rangle=(76.6\pm1.0)\%$. We implement this protocol using photons at 1550 nm, making it well suited for quantum repeaters, where it could enable more efficient entanglement distribution. Beyond quantum communication, it could also enhance fusion-based quantum computing.

\end{abstract}

\maketitle

%%%%%%%%%%%%%%%%%%%%%%%%%%%%%%%%%%
%%%%%%%%    Main text    %%%%%%%%%
%%%%%%%%%%%%%%%%%%%%%%%%%%%%%%%%%%
%____________________________________________________________________________
\section{Introduction}
%____________________________________________________________________________
Quantum networks rely on distributing entanglement between distant parties. Entanglement swapping, originally proposed as a way to entangle particles that have never interacted~\cite{Zukowski1993} and first demonstrated with photons~\cite{pan_experimental_1998}, is a key primitive for this task. Its role extends to several applications, including quantum key distribution~\cite{Braunstein2011, lo2012, Yin2016} and quantum repeaters~\cite{Briegel1998, pu2021, Azuma2023, Krutyanskiy2023}, and it is closely related to quantum teleportation~\cite{Bennet1993, Braunstein1995, daurelio_boosted_2025}.
Beyond communication, entanglement swapping via Bell-state (fusion) measurements can also be used to generate large entangled resource states from smaller building blocks~\cite{browne2005, bartolucci2023, Hauser2025}. These fusion operations are the basis of linear-optical quantum computing~\cite{Knill2001} and photonic quantum information processing, including measurement-based and distributed quantum protocols~\cite{Raussendorf2001, raussendorf2003, lim2005, browne2005, bartolucci2023}. 

The key component of entanglement swapping is a Bell-state measurement (BSM)~\cite{pan_experimental_1998}. Linear-optical implementations of this measurement have been demonstrated with bulk optics~\cite{Weinfurter1994, Mattle96, Michler96}, are compatible with a broad range of wavelengths, and can be directly translated into photonic integrated circuits~\cite{Wang2020}. However, there is a fundamental theoretical bound on the success probability of these measurements with only linear-optical elements: BSMs yield ambiguous outcomes in half of the cases~\cite{Luetkenhaus1999,calsamiglia2001}, limiting the success probability to 50\%. This strongly restricts their usability in scalable quantum network architectures. Higher success probabilities can, in principle, be achieved in photonic platforms by using hyperentangled states~\cite{Kwiat1998, Walborn2003,Schuck2006, Barbieri2007, Li2017,Williams2017}, nonlinear optical elements~\cite{Kim2001,akin_faithful_2025,tsujimoto_experimental_2025} or atomic- or matter-based systems~\cite{welte_nondestructive_2021,nielsen_photon_2026}. These methods can enable deterministic BSMs, but typically require more complex interfaces to photonic channels.

Ancillary photons offer a way past the $50\%$ bound that keeps the advantages of linear optics. By sending additional photons into the measurement setup, linear-optical BSMs with success probabilities arbitrarily close to unity become possible~\cite{grice2011, ewert2014, olivo2018, Bianchi2026}. Such \textit{boosted} BSMs have recently been demonstrated experimentally~\cite{bayerbach2023} and applied to quantum teleportation~\cite{daurelio_boosted_2025} and photonic quantum computation~\cite{Hauser2025, guo_boosted_2026}. 
In quantum communication protocols over lossy channels, inefficient BSMs at intermediate nodes directly reduce the entanglement distribution rate. In such schemes, the ancillary photons are generated locally at the BSM node and therefore do not suffer from channel losses. Boosting the BSM in architectures such as all-photonic quantum repeaters~\cite{azuma_all_photonic_2015,Hasegawa2019} or measurement-device-independent quantum key distribution~\cite{lo2012} can therefore increase the achievable entanglement distribution rate and extend the usable transmission range~\cite{azuma_all_photonic_2015}. 

In this work, we demonstrate boosted entanglement swapping using a boosted Bell-state measurement. Two photonic Bell states are generated, and one photon from each pair is jointly measured in the boosted BSM using ancillary photons, achieving an average fidelity of the swapped states of $\langle F^{\mathrm{b}}\rangle=(76.6\pm1.0)\%$. This enables successful entanglement swapping between the two remaining photons, which we verify through entanglement witnesses and full quantum state tomography.
\begin{figure}[b]
    \centering
    \includegraphics[width=\linewidth]{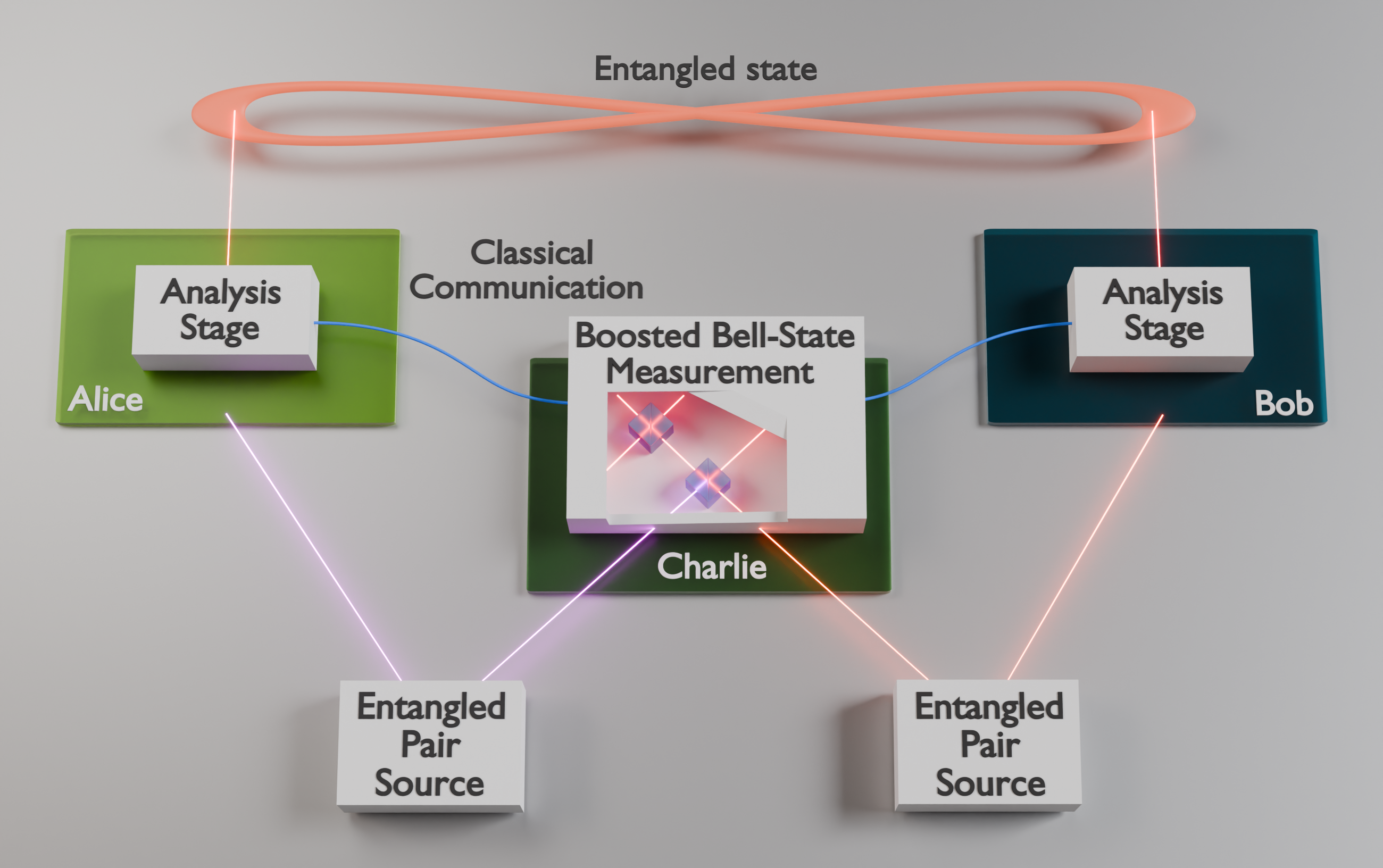}
    \caption{\textbf{Boosted entanglement swapping.} Two independently generated photon pairs ($\ket{\Psi^-}$) are shared between the parties Alice and Charlie and between Bob and Charlie. Charlie performs a boosted BSM and announces the outcome to the communicating parties. This measurement identifies which Bell state is shared by Alice and Bob.
    }
    \label{fig:concept_figure}
\end{figure}
\begin{figure*}
    \centering
    \includegraphics[width=0.90\linewidth]{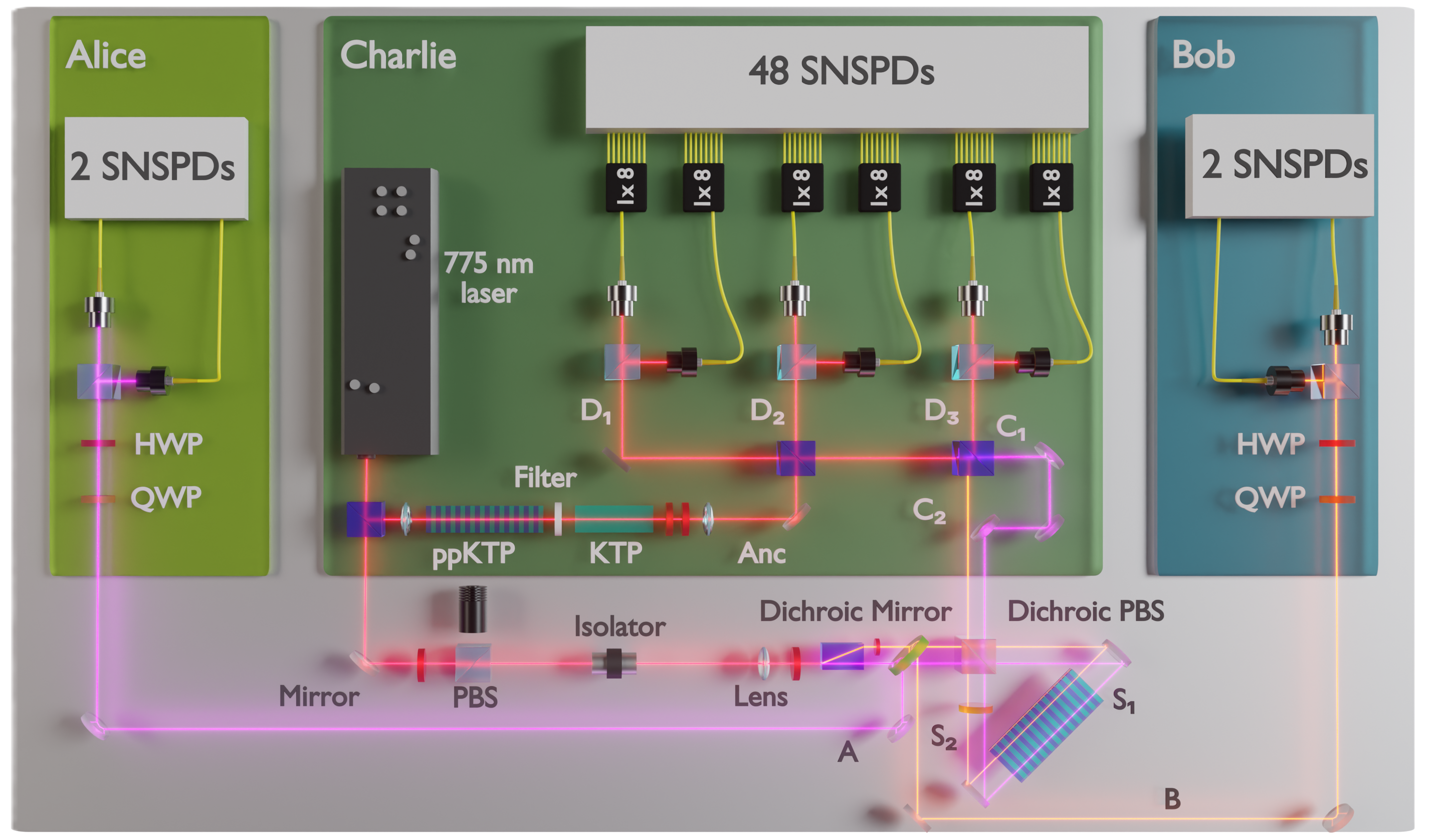}
    \caption{\textbf{Setup.} Both Bell pairs ($S_1$, $S_2$) are generated in a double-loop Sagnac interferometer (bottom). The ancillary state (Anc) is generated in a separate ppKTP crystal inside Charlie's node. In this source, a KTP crystal is used to compensate the walkoff between polarisations introduced in the ppKTP crystal. \textbf{Charlie.} Implementation of the boosted BSM. Photons arriving from different entangled pairs ($C_1$ and $C_2$) are sent into the BSM, consisting of two beam splitters. Each output mode $D_1$\textendash$D_3$ is split by a PBS, and each PBS output is evenly split by 1$\times$8 splitters into eight SNSPDs for pseudo-photon-number resolution. \textbf{Alice and Bob.} Polarisation analysis stages (QWP, HWP, PBS) for quantum-state tomography. Each of the parties receives one photon from either of the generated Bell pairs, labelled $A$ and $B$. Based on Charlie's BSM outcome, they perform quantum-state tomography on their shared state, to validate the correct operation of the entanglement swapping scheme.} 
    \label{fig:setup}
\end{figure*}
\section{Results}
\subsection*{Theory background}
%____________________________________________________________________________
We aim to distribute entanglement between two parties (Alice and Bob) via entanglement swapping. A third party (Charlie) shares an entangled state with both Alice and Bob (see Fig.~\ref{fig:concept_figure}). The two-qubit entangled state to be used in the protocol is the Bell state:
\begin{equation}
\ket{\Psi^-}=\frac{1}{\sqrt{2}}\left(\ket{01} - \ket{10} \right),
\end{equation}
where $\ket{0}$ and $\ket{1}$ denote the logical states. The joint state $\ket{\Psi^-}_{AC_1} \otimes \ket{\Psi^-}_{BC_2}$ can be written as:

\begin{equation}
\label{eq:entswap}
\begin{aligned}
    \ket{\Psi^-}_{AC_1}& \otimes  \ket{\Psi^-}_{BC_2} = \\ \frac{1}{2}\bigg(& \ket{\Psi^-}_{AB}\ket{\Psi^-}_{C_{12}} -\ket{\Psi^+}_{AB}\ket{\Psi^+}_{C_{12}}\\+&\ket{\Phi^+}_{AB}\ket{\Phi^+}_{C_{12}} - \ket{\Phi^-}_{AB}\ket{\Phi^-}_{C_{12}}\bigg),
\end{aligned}
\end{equation}
where the subscripts refer to Alice (A), Bob (B), and Charlie (C) and $\ket{\Psi^\pm}$, $\ket{\Phi^\pm}$ are the four Bell states.
A BSM by Charlie projects Alice's and Bob's qubits onto the corresponding Bell state. Once Charlie announces the result via classical communication, Alice and Bob know which entangled state they share.

In photonic implementations, the BSM that Charlie conducts is typically performed by letting the two photons interfere at a balanced beam splitter (BS). The outcome of the BSM is then assessed by analysing the distribution of the photons in the output of the BS, projecting in the degree of freedom carrying the entanglement.
However, only two of the four possible Bell states will produce unique detection patterns, while the others will appear ambiguous. For these measurement events, the local correction to be applied is not defined; therefore, the measurement result is considered a failure. We refer to the theoretical probability of an unambiguous BSM outcome as the success probability ($50 \%$ for the standard BSM, $62.5 \%$ for the boosted scheme used here). A related but different value is the measured fraction of detected events that are assigned to an unambiguous outcome, which is the acceptance rate $p_{\mathrm{a}}$ and is defined as:
\begin{equation}
\label{eq:accrate}
    p_{\mathrm{a}}=\frac{N_{\mathrm{accepted}}}{N_{\mathrm{accepted}}+N_{\mathrm{ambiguous}}}.
\end{equation}
Here $N_{\mathrm{accepted}}$ and $N_{\mathrm{ambiguous}}$ are the numbers of unambiguous and ambiguous events, respectively. Because experimental imperfections can cause events to be classified as unambiguous, $p_{\mathrm{a}}$ can exceed the success probability.
To surpass the 50\% limit on the success probability, we use a scheme that is based on a two-photon ancillary state $\ket{\nu}$. This method, described in~\cite{ewert2014} and experimentally realised in~\cite{bayerbach2023, Hauser2025, daurelio_boosted_2025}, raises the theoretical success probability to $62.5\%$. To implement the boosted BSM, one requires the ancillary two-photon state $\ket\nu=\frac{1}{\sqrt{2}}(\ket{2_H,0_V}+\ket{0_H,2_V})$, with the two photons occupying the same spatial mode. In this notation $\ket{n_H,m_V}$ represents the Fock state with $n$ and $m$ photons in the horizontal and vertical polarisation. This ancilla interferes with one output of the BS of a standard BSM at a second BS (see Fig.~\ref{fig:setup}).

To prove that there is entanglement between the two parties, an entanglement witness can be measured, as defined in~\cite{Guehne2002,Barbieri2003}. The entanglement witness is an operator, here denoted $\hat{W}$, that has a negative expected value for non-separable states and non-negative for separable ones. For a Werner state $\rho=q\ket{\Psi^-}\bra{\Psi^-}+\frac{1-q}{4}I$, this operator has a simple representation. In the polarisation encoding ($\ket{0}\equiv \ket{H}$, $\ket{1} \equiv \ket{V}$), used in the experiment, the operator reads~\cite{Barbieri2003}:
\begin{equation}
\label{eq:witness}
\begin{aligned}
\hat{W} = \frac{1}{2}\big(&|HH\rangle\langle HH| + |VV\rangle\langle VV|\\
+ &|DD\rangle\langle DD| 
+ |AA\rangle\langle AA| \\ -&|LR\rangle\langle LR|
- |RL\rangle\langle RL|\big),
\end{aligned}
\end{equation}
where $\ket{D/A} = \frac{1}{\sqrt{2}}\big(\ket{H}\pm\ket{V}\big)$, $\ket{L/R} = \frac{1}{\sqrt{2}}\big(\ket{H}\pm i\ket{V}\big)$ and the mode subscript A/B is omitted.

For the remaining Bell states, the witness is obtained by applying the corresponding local Pauli rotation to $\hat{W}$.
%____________________________________________________________________________

%____________________________________________________________________________
\subsection*{Experimental setup}
The experimental implementation of the protocol is depicted in Fig.~\ref{fig:setup}. Both Bell pairs are generated via spontaneous parametric down-conversion (SPDC) in a periodically poled potassium titanyl phosphate (ppKTP) crystal. The crystal is placed in a Sagnac interferometer with a double-loop geometry to generate both pairs within the same crystal~\cite{daurelio_boosted_2025}. A pulsed laser sends 775~nm picosecond pulses into the crystal, generating photon pairs at 1550~nm, in the telecom C-band. The generated photons are coupled into optical single-mode fibres and spectrally filtered using $1.5$ nm filters. From each generated pair, one photon is sent to Charlie (labelled  $C_1$ and  $C_2$ in Fig.~\ref{fig:setup}), while the other two (labelled  $A$ and  $B$) are transmitted to Alice and Bob. The two Bell-state sources are configured to generate the state $\ket{\Psi^-}$, with fidelities of $F_{\mathrm{S_1}}=(97.08\pm0.04)\%$ for source 1 ($S_1$) and $F_{\mathrm{S_2}}=(97.09\pm0.04)\%$ for source 2 ($S_2$), see Fig.~\ref{fig:setup}.

Charlie's node performs the boosted BSM. The ancillary photons are generated in a state $\ket{\nu_0}=\ket{1_H,1_V}$ also via SPDC, as depicted in Fig.~\ref{fig:setup}. By inserting a half-wave plate with the fast axis rotated by $22.5^\circ$, the state becomes:
\begin{equation}
    \ket{\nu_p}=\frac{1}{\sqrt{2}}(\ket{2_H,0_V}-\ket{0_H,2_V}),
\end{equation}
 which is equivalent, up to a phase between $H$ and $V$, to $\ket{\nu}$ in polarisation encoding. To perform the boosted BSM, the photons received by Charlie are sent into the two inputs of a balanced beam splitter (Charlie's node in Fig.~\ref{fig:setup}). One of its outputs is sent into the input of a second balanced beam splitter while the ancillary state enters the other input, labelled $Anc$. The three output modes $D_1$, $D_2$, and $D_3$ of the two beam splitters are then analysed in polarisation and photon number. Each of the two output modes of every polarising beam splitter (PBS) is sent into a 1$\times$8 fibre splitter, whose outputs are connected to eight independent superconducting nanowire single-photon detectors (SNSPDs), thereby providing pseudo-photon-number resolution (pseudo-PNR). Each event's click pattern is assigned to one of five classes (the four Bell states or an ambiguous outcome) according to the ideal detection patterns of the scheme~\cite{ewert2014,bayerbach2023}. On Alice's and Bob's sides, a set of quarter-wave plates (QWP) and half-wave plates (HWP) enables projective measurements in any combination of the Pauli bases $\{X, Y, Z\}$.
\begin{figure*}
    %\centering
    \includegraphics[width=\textwidth]{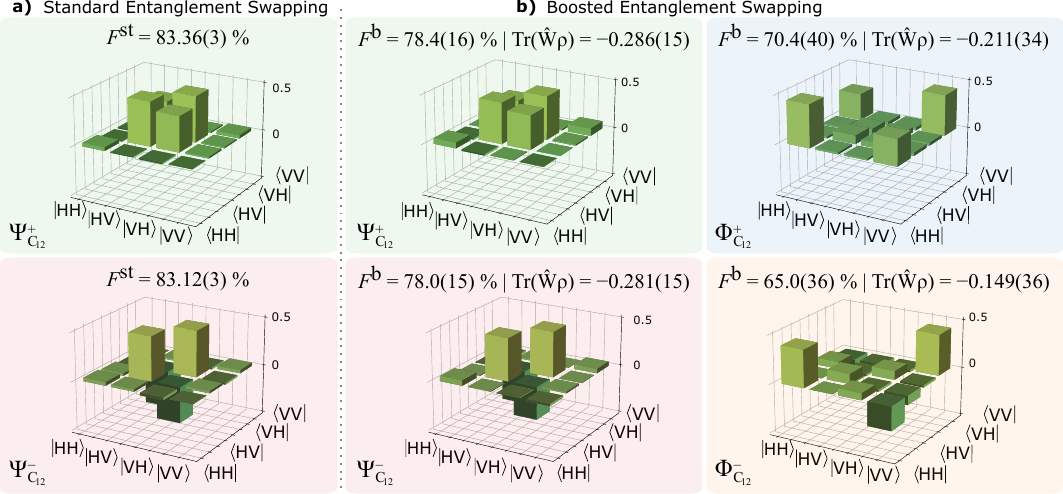}
    \caption{\textbf{Density matrices for the generated entangled states.}
    Each box shows the reconstructed density matrix $\mathrm{Re}(\rho)$ of the resulting swapped state for a specific BSM outcome, indicated in the bottom left corner. The imaginary part of the density matrix is not shown. \textbf{a)} Density matrices of the swapped states obtained with the standard BSM. The measurements for which the BSM projected onto $\ket{\Phi^\pm_{C_{12}}}$ are not shown, as they are ambiguous. \textbf{b)} Density matrices for the boosted entanglement swapping scheme. In this case the BSM can project onto all four Bell states.} 
    \label{fig:density_matrices}
\end{figure*}
%____________________________________________________________________________
\subsection*{Measurement results} \label{Measurement results}
%____________________________________________________________________________
Polarisation measurements are performed by both Alice and Bob to characterise the shared quantum state, enabling quantum-state tomography. The results are sorted according to the BSM outcome. We measure both the standard and boosted BSM by recording both four-fold (standard BSM) and six-fold (boosted BSM) coincidence events simultaneously. A maximum likelihood estimation (MLE) method is used to reconstruct the six density matrices and the corresponding results are shown in Fig.~\ref{fig:density_matrices}. We obtain state fidelities of the swapped entangled state of $F^{\mathrm{st}}_{\Psi^+}=(83.36\pm0.03)\%$ and $F^{\mathrm{st}}_{\Psi^-}=(83.12\pm0.03)\%$, averaging to $\langle F^{\mathrm{st}}\rangle=(83.23\pm0.02)\%$ for the standard BSM, as shown in Fig.~\ref{fig:density_matrices} $\mathbf{a)}$. The uncertainties are obtained by repeating the MLE on simulated data sets drawn from Poisson distributions centred on the measured counts. They therefore account for statistical (counting) errors only; systematic errors, e.g. from wave-plate miscalibration or drifts in the setup, are not included. 
The acceptance rate for this measurement is $p^{\mathrm{st}}_\mathrm{a}=(51.50\pm0.01) \%$. The acceptance rate exceeds the theoretical success probability of $50 \%$ because of experimental imperfections, such as higher-order emissions and imperfect beam-splitter ratios.

For the boosted BSM, shown in Fig.~\ref{fig:density_matrices}~$\mathbf{b)}$, we obtain an acceptance rate of $p^{\mathrm{b}}_\mathrm{a}=(69.39\pm0.39)\%$. The fidelities of the four swapped states are $F^{\mathrm{b}}_{\Psi^+}=(78.4\pm1.6)\%$, $F^{\mathrm{b}}_{\Psi^-}=(78.0\pm1.5)\%$, $F^{\mathrm{b}}_{\Phi^+}=(70.4\pm4.0)\%$ and $F^{\mathrm{b}}_{\Phi^-}=(65.0\pm3.6)\%$, resulting in a weighted average fidelity of the swapped states of $\langle F^{\mathrm{b}} \rangle=(76.6\pm1.0)\%$. Projections onto $\ket{\Phi^\pm}$ herald swapped states with lower fidelities than projections onto $\ket{\Psi^\pm}$. This difference arises from an asymmetry of the scheme: $\ket{\Psi^+}$ and $\ket{\Psi^-}$ are already distinguished after the first BS, whereas distinguishing $\ket{\Phi^+}$ from $\ket{\Phi^-}$ additionally requires
interference with the ancillary photons at the second BS. Projections onto $\ket{\Phi^\pm}$ are therefore more sensitive to imperfect indistinguishability between ancillary and input photons and to higher-order emissions.

%%%%%%%%%%%%%%%%%%%%%%%%
Although the theoretical limit of the scheme is $62.5\%$, imperfections in the setup can increase the recorded acceptance rate, generally at the price of a reduced fidelity. These effects can be identified as contributions from higher-order emissions and imperfect BSs and PBSs. Another source of artificial increase in the acceptance rate can be imbalances in the pseudo-PNR part of the measurement. In general, these do not necessarily reduce the fidelity of the states, but by sampling different total photon-number states with different probabilities, they can skew the measured distribution. The result is a modified acceptance rate, but in this case the cost is an effective reduction of the total events. This shows that acceptance rate alone is not sufficient to characterise a BSM, but should be assessed together with the fidelity of the resulting states.

%%%%%%%%%%%%%%%%%%%%
We evaluate the entanglement witness of Eq.~(\ref{eq:witness})~\cite{Guehne2002, Barbieri2003} directly from the measured coincidence counts, without relying on the state reconstruction. A value of $\mathrm{Tr}(\hat{W}\rho) < 0$ is a sufficient criterion for certifying entanglement, where $\rho$ represents the density matrix and $\hat{W}$ is the operator described in Eq.~(\ref{eq:witness}). For all four boosted BSM outcomes, the swapped states show a negative measured witness value, with an average value of $\langle\mathrm{Tr}(\hat W \rho)\rangle=-0.268\pm0.010$.
%____________________________________________________________________________
\section{Discussion and Conclusion}
%____________________________________________________________________________
In this work, we demonstrate an implementation of an entanglement swapping protocol featuring a boosted BSM. Here, the use of ancillary photons increases the number of accepted events, raising the acceptance rate from $p^{\mathrm{st}}_\mathrm{a}=(51.50\pm0.01) \%$ to $p_{\mathrm{a}}^{\mathrm{b}}=(69.39\pm0.39)\%$. We show that every one of the four Bell-state outcomes of the BSM heralds entanglement between photons that never interacted. Quantum-state tomography reveals the complete swapped state for all four BSM outcomes, resulting in an average fidelity for the boosted BSM of $\langle F^{\mathrm{b}}\rangle =(76.6\pm1.0)\%$, starting from $\langle F^{\mathrm{st}}\rangle=(83.23\pm0.02)\%$ for the standard BSM. An entanglement witness evaluated directly from the coincidence counts certifies that Alice and Bob share an entangled state for every outcome.

Our system relies only on linear-optical components to perform the projective Bell-state measurement, requiring no nonlinear or matter-based elements. In addition, it is adaptable in principle to a wide range of wavelengths and bandwidths. A possible application would be a boosted quantum repeater scheme. There, a repeater node performs a Bell-state measurement to establish entanglement between two distant parties. Boosting this measurement increases the probability of successful entanglement swapping at the repeater node for a given channel transmission. Depending on the repeater architecture and the channel losses, the resulting increase in distribution rate could enable operation over longer distances~\cite{azuma_all_photonic_2015, li2019}. Furthermore, the additional resources (ancillary photons and photon-number-resolving detection) are required only at the repeater node, not at the communicating parties.
The chosen wavelength in the telecom C-band ensures excellent compatibility with existing fibre networks and also opens the possibility of integrating the system on silicon photonic circuits~\cite{alexander_manufacturable_2025}.
%\clearpage
\begin{acknowledgments}
We acknowledge the support from the Federal Ministry of Research, Technology and Space (BMFTR, projects SiSiQ: FKZ 01N14920, PhotonQ: FKZ 01N121124A, QR.N: FKZ 16KIS2207, TD.QR: FKZ 16KISS026, QCyber: 16KIS2590K), and the Deutsche Forschungsgemeinschaft (DFG, German Research Foundation, 431314977/GRK2642, 516238647/SFB 1667, 563437379/SPP2514), the Carl Zeiss Foundation, the Centre for Integrated Quantum Science and Technology (IQST), the Ministry of Science, Research and Arts Baden-Württemberg via the IQST Graduate School @QuantumBW.
\end{acknowledgments}
%\bibliography{refs}

%apsrev4-2.bst 2019-01-14 (MD) hand-edited version of apsrev4-1.bst
%Control: key (0)
%Control: author (8) initials jnrlst
%Control: editor formatted (1) identically to author
%Control: production of article title (0) allowed
%Control: page (0) single
%Control: year (1) truncated
%Control: production of eprint (0) enabled
%

\begin{footnotesize}
\noindent

\end{footnotesize}

\end{document}